\documentclass[prc,twocolumn,floatfix,groupedaddress,showpacs,superscriptaddress] {revtex4}
\usepackage{graphicx}
\usepackage{dcolumn}
\usepackage{mathrsfs}
\usepackage{bm}
\usepackage{dsfont}
\usepackage{dcolumn}
\usepackage[usenames]{color}
\usepackage{amssymb}
\usepackage{amsmath} 
\usepackage{makeidx}
\usepackage{array}
\usepackage{multirow}
\usepackage{bm} 
\usepackage{soul} 
\usepackage[normalem]{ulem}
\DeclareMathAlphabet{\mathpzc}{OT1}{pzc}{m}{it} 

\usepackage{color}

\definecolor{OliveGreen}{rgb}{0,0.6,0}

\begin{document}
\title{Experimental study of the relative probability of high-spin isomeric states population in ($\alpha$,n)-reactions}

\author{T. V. Chuvilskaya}
\affiliation{Skobeltsyn Institute of Nuclear Physics, Lomonosov Moscow State University,
119991 Moscow, Russia}
\date{\today}

\begin{abstract}
Results of the investigations of the yield of high-spin and low-spin isomers in reactions   $^{41}$K($\alpha $,n)$^{44}$Sc, $^{86}$Sr ($\alpha $,n)$^{89}$Zr, $^{112}$Sn($\alpha$,n)$^{115}$Te, and $^{134}$Ba($\alpha$,n)$^{137}$Ce in the energy range of the alpha particles 15 $\div$ 31 MeV based on off-beam measurements of induced activity of members of the isomeric pair are presented. The anomalous behavior of the isomeric cross-section ratios for the first of these reactions is  confirmed. Uniquely large isomeric cross-section ratios for the second and third ones are obtained. The features of the fourth reaction, which are promising for its application in fundamental research, are revealed.  Quality of the description of this dynamics by popular computer codes is analyzed. 
\end{abstract}

\maketitle

\section{Introduction}
Nuclear isomers i. e. excited metastable states of nuclei play an important role in a lot of areas of nuclear studies and fields of application of nuclear science. In a large number of papers and many monographs (see, for example \cite{sieg}) the phenomenon of nuclear isomerism is studied, analyzed and discussed. The study of manifestations of this phenomenon served as one of the substantiations of the various nuclear models, and made a significant contribution to the theory of multipole $\gamma$-radiation. The existence of long-living isomers is associated with a strong decrease in the probability of gamma transitions. There are three known reasons for the appearance of the long-living (metastable) states:

\noindent
1. The spin of a metastable state differs from the spins of all states lying below by 3 or more units $\hbar$ (spin-gap isomers).

\noindent
2. Despite the presence of lower-lying levels with spins close to the spin of the isomeric state, all such levels differ greatly from it in the value of $K$ projection of the spin onto the axis of symmetry of the nucleus ($K$-isomers).

\noindent
3. An isomeric state of the nucleus differs sharply in shape  from the lower-lying ones  (shape isomers or fission isomers in heavy nuclei).

In the studies presented in this paper the examples of states belonging to the first type are considered. 

The range of research  and practical applications of well-studied nuclear isomers is also wide. They are often used as tools for studying nuclear structure (see, for example Ref. \cite{struct}), fundamental symmetry violation \cite{mu,ch0,kurg,ald}, and astrophysical  nuclear synthesis processes \cite{astro1,no}. 

The isomers are widely used in various  areas of applied science, radiochemistry, activation analysis, radiopharmacology, immunology. The matter is that the radiation of a nuclear isomer is a good indicator of the presence of a chemical element in a chemical, geological or biological sample; a nuclear process that has occurred under laboratory or natural conditions, as well as of characteristics of such processes. Isomers are high-capacity energy accumulators. Possibilities are being discussed and experiments are being performed to develop a method for using this energy \cite{tkalya,karam}.  At the present time,  further development of the investigations of the properties of nuclear isomers is expected in the context of unstable isomeric beams production \cite{rings}. That is why various studies of the properties of isomeric states, methods of producing them, and the processes involving such states are popular and promising.

Isomeric states population and production of them in the amount required for research and applied purposes is one of the basic problems  of the just mentioned studies. Selection of a reaction suitable for obtaining a certain isomer in a significant amount with an acceptable percentage of impurities is a delicate task. Neutron, $\gamma$-quanta, light-. and heavy-ion beams of various energies as well as fission processes are used for these purposes.

In the case that ordinary spin-gap isomers are considered, the ratio of cross sections of producing of a certain pair of states (high-spin and low-spin) in one and the same residual nucleus allows one to obtain an important information about angular momentum dynamics of a preceding reaction and spin dependence of nuclear level density. The discussed dynamics depends on the properties of a target, projectile, and emitted  particles. Search for reaction conditions being optimal to produce any isomer and especially high-spin one seems to be a vital issue. 

Nuclear reactions induced by $\alpha$-particles of energy several MeV/A figure prominently in low-energy nuclear physics. They are a tool for studying the structure of nuclei, features of interaction of composite particles, as well as a tool for populating various nuclear states. These reactions play a prominent role in astrophysical processes \cite{astro2}. They are used to obtain radioactive sources used in  medicine  \cite{med,aik1,aik2}, including radionuclide therapy, immunology, etc. Among radiopharmaceutical products, produced by such reactions there are also those that contain nuclear isomers, in particular $^{89m}$Zr \cite{aik3}. 

Alpha particle beam possesses unique properties extremely important for populating high-spin states. It carries a much higher angular momentum in comparison with lighter particles at one and the same excitation energy of the compound nucleus, and the potential barrier between this particle and the target nucleus is significantly lower than the barrier typical for reactions with heavy ions. These properties are discussed below. This circumstance is, to a large extent, the motivation for this work and our other studies.

The values of isomer formation cross sections for typical $\alpha $-induced nuclear reactions such as ($\alpha $,n), ($\alpha $,2n), etc. are presented in many papers, see, for example Refs. \cite{br,gr,ne,av,av2,ba4,ba5,ba3,ba2,vi,was0,was,gl1,ch1,tu,par,na,su,ch2,ch3,ch4}.

In the present paper the results of the investigations of the energy dependence of the yield of a high-spin isomeric state relative to the yield of a low-spin one in one and the same residual nucleus for reactions   $^{41}$K($\alpha $,n)$^{44}$Sc, $^{86}$Sr ($\alpha $,n)$^{89}$Zr, $^{112}$Sn($\alpha$,n)$^{115}$Te, and $^{134}$Ba($\alpha$,n)$^{137}$Ce in the energy range of the $\alpha$-particles E=15 $ \div$ 31 MeV are presented.  The measure of the relative isomeric yield -- so-called isomeric cross-section ratio (ICSR) or, in short, the isomeric ratio --  was studied experimentally by means of off-beam measurements of the induced activity of members of an isomeric pair. The activation method is a reliable tool for identification of residual isotopes. This is its advantage over the method of direct  in-beam measurements of the excitation functions by detecting of the reaction products. In addition, measuring the relative yields of isomers in one and the same experiment often makes it possible to exclude some sources of systematic errors due to their direct reduction. All measurements of the yields were performed using the alpha-particle beam of the SINP MSU cyclotron. Induced activity was measured using  $\gamma$-spectrometers based on Ge(Li)-detectors. In the present paper the results of latest measurements and/or the ones improved through the handling of the activation data with the use of the optimal extraction formula from \cite{va} are demonstrated. These data are unpublished or differ from the ones presented in databases, including those that are placed in the databases with a link to the work of our group. Some of the results are presented in the literature only in the form of figures, therefore the current paper contains the tables of the values of ICSR convenient to use in the databases.  

In general, ICSR are the values, strongly dependent on the properties of the processes occurring at the moment of collision, as well as after it in the compound and residual nucleus. They are significantly more sensitive to details of these processes compared to the total excitation functions of nuclear reactions and angular distributions of primary reaction products. 

A hope to reveal the details of the discussed processes  determining the value of some ICSR arises only when using advanced codes for calculating the cross sections of nuclear reactions. Recently the analysis of the excitation functions is based on large-scale data-containing codes, for example,  the EMPIRE 3.1 \cite{he2} and the TALYS  \cite{ko}. Codes of such a type allow one to take into account a large number of characteristics of nuclear reactions and nuclei involved in them. It is these codes that were used in this work in attempts to describe the experimental data. As a result of the comparison of experimental and theoretical data, presented both in this and in other works, a variegated pattern emerges. It is far from always possible to describe the experimentally obtained dependence of ICSR on energy.

\section{A beam of alpha particles as a unique means for obtaining high-spin isomers.}

\begin{figure}
\includegraphics[width=1.30\linewidth]{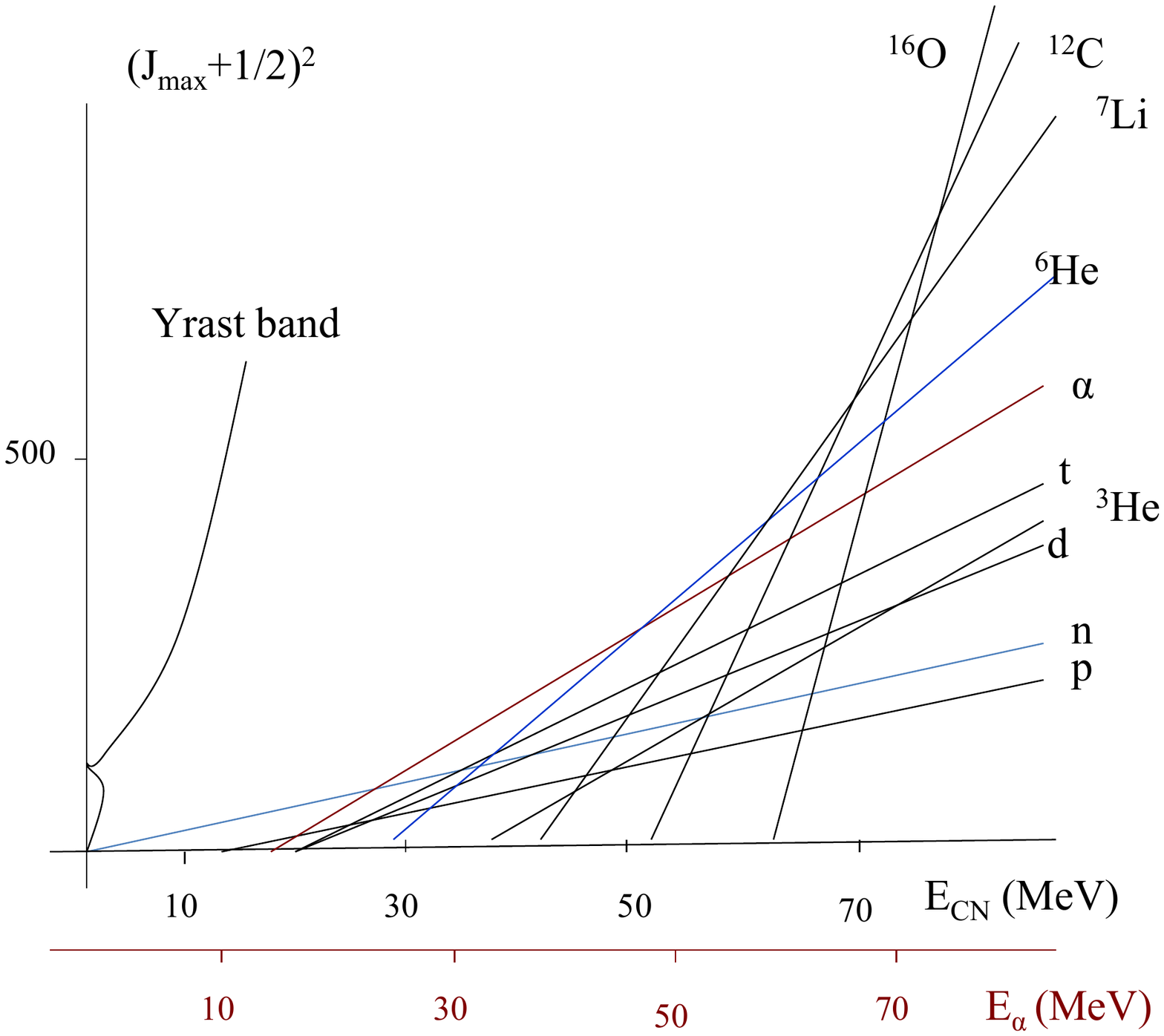} 
\caption{Maximal angular momentum contributed by various particles into 
the  $^{120}$Sn compound nucleus.\label{fig1}}
\end{figure}

Let us point out, first of all, the specific properties of a beam of low-energy alpha particles that make this beam especially interesting for obtaining and studying high-spin isomeric states. The  matter is that in the process of collision with a target nucleus, the alpha particle brings a large angular momentum into the compound nucleus.  Figure \ref{fig1} shows the results of a simple estimate of the maximum angular momentum $J_{max}$ brought by various particles into the compound nucleus $^{120}$Sn, as a function of on the excitation energy of the nucleus $E_{CN}$. The mechanism of complete fusion is assumed. Expression $(J_{max}+1/2)^2=2 \mu (E_{CN}-E_0)(R_t+R_i+2fm)^2/\hbar^2$, where $\mu$ is the reduced mass of the entrance channel, $E_0$ -- the fusion energy of the incident particle and the target nucleus, $(E_{CN}-E_0)$ -- the kinetic energy of the incident particle in the c. m. system,  $R_t$ and $R_i$ are the radii of the corresponding nucleus and particle, is used for the estimate. The lower scale shows the energy of the bombarding alpha particle, corresponding to the energy of the resulting compound nucleus. For relatively large $J_{max}$, the average angular momentum brought by the incident particle is approximately (2/3)$J_{max}$. As can be seen from the figure, the $\alpha$-particle in the excitation energy range from 25 to 50 MeV brings the largest angular momentum into the compound nucleus in comparison with other particles and heavy ions.

Evidently this should result in a higher probability of excitation of high-spin states of a residual nucleus  in reactions induced by $\alpha$-particles compared to other reactions. So, these reactions should lead, as a rule, to a relatively large formation of high-spin isomers in the subsequent  $\gamma$-cascade process. 

\section{The scheme of ICSR measurements. Experimental setup and method of processing the obtained experimental results}

All measurements of the isomeric yields were performed in different years using the $\alpha$-particle beam of the SINP MSU cyclotron  and the $\gamma$-spectrometers with a Ge (Li)-detectors. The targets were irradiated with $\alpha$-particles with the energy 25 or 31 MeV and the beam current F $\simeq$ 2 $\mu$A. Targets enriched with the required isotope were used in all these experiments.

Irradiation of each sample in a vacuum chamber was performed with a cyclotron beam after focusing it with quadrupole lenses. Powdered materials or foils of the investigated substances were wrapped in aluminum foil. After irradiation, the target was quickly locked out of the chamber and, behind the shield, was removed from the adapter. After that the target moved to the room where the spectrometer is located. A radiated target was fixed on a frame and placed at distance 5 $\div$ 10 cm from the Ge(Li) detector. The choice of a specific distance was dictated by restrictions on the detector loading. When measuring the relative yield of isomers, as a rule, the most intense $\gamma$-lines belonging to each of the isomeric states were selected. This method allows one to study isomers with a half-life of about $t\geq$ 1 min.

Various Ge(Li) detectors were integrated into the scheme in different experiments. They had a sensitive volume of 3, 10 or 50 cm$^3$ and the resolution of the $\gamma$-spectrometer at $E_{\gamma}$ = 1.3 MeV was, respectively, 3.5, 4.5, and 6.0 keV. The tract consists of a preliminary charge-sensitive amplifier, the first stage of which, mainly determining the system noise, is assembled on three parallel-connected field-effect transistors. Then the signal is fed to a linear amplifier with shaping circuits and a cathode follower. The amplifier is connected to an amplitude analyzer. The $\gamma$-spectrometer was calibrated in terms of energies and relative efficiencies using standard calibration sources $^{207}$Bi, $^{203}$Hg,$^{ 60}$Co, $^{22}$Na, $^{152-154}$Eu, $^{137}$Cs, $^{54}$Mn, $^{56}$Co, $^{88}$Y. The error in determining the relative efficiency is 5\%. The linearity of the energy scale of the $\gamma$-spectrometer was observed to be $\pm$ 1 keV in the energy range 200 $\div$ 4000 keV.

Ratio of the cross sections for the formation of high-spin and low-spin isomers $(\sigma_m/\sigma_g)$ (below in all cases this definition of the ratio is considered regardless of which of the states is located higher in energy) was determined from the value of the ratio of the induced activity of these isomers. Measurements of such a type are possible, naturally, if both of these states of the studied isotope are unstable and have suitable  half-lives for measurements.

Independent experimental measurements of the activities of the high-spin and low-spin states for determining $(\sigma_m/\sigma_g)$ ratio are less accurate than the determination of this ratio in one and the same experiment from the measured $\gamma$-spectrum of one and the same sample, therefore in the discussed experiments the ratio of the cross sections $(\sigma_m/\sigma_g)$ was determined by this means. When measuring the $\gamma$-spectrum, the spectrometer was calibrated in energy at the beginning and at the end of measurements using standard sources. The residual activity was measured, which made it possible to exclude the lines related to the background activity from the spectra.

The $\gamma$-lines belonging to only one of the isomers were reliably distinguished from the rest of spectrum. In the case that  a full-energy peak contains a contribution from the spectrum of another isomer, then it was taken into account. The belonging of these  $\gamma$-lines to the investigated isomer was determined from the energy of  $\gamma$-transitions and the half-life \cite{fi}. Since at present the level diagrams for many nuclei have been reliably established up to an energy of 1.5 $\div$ 2.0 MeV, we used tabular data to identify the $\gamma$-lines. The $\lambda$ values, the relative intensities $I_\gamma$ and the coefficients of internal conversion $\alpha_n$ were taken from Refs. \cite{led,reus}. Nevertheless, for each radioactive isotope, the decrease in the  $\gamma$-line intensities with time was measured and the constants of the corresponding decays $\lambda_m$ and $\lambda_g$ were determined. For the examples presented in this work, no contradictions between the data were found. By summing the pulses under the full-energy peaks of the total absorption of the $\gamma$-lines, the areas $S_m$ and $S_g$ related to the the high-spin and low-spin states of the isotope were determined.

The relative yield of isomers ${P_m }/{P_g }$  in  a certain reaction was the subject of the study. Its value is proportional to the ratio ${S_m }/{S_g }$ of the areas of $\gamma$-peaks corresponding to the high-spin and low-spin isomeric states of the nucleus under study. Thus, taking into account the detector efficiency and the percentage of $\gamma$-line decays (the probability of the $\gamma$-transition) the relative yield is expressed through the ratio of the sample activities. In the absence of an isomeric $\gamma$-transition the formula takes the form,

\begin{equation}
\frac{S_m }{S_g } = \frac{{P_m \lambda _g \left( {1 - e^{ - \lambda mt_0 } } \right)e^{ - \lambda mt_1 } \left( {1 - e^{ - \lambda mt_2 } } \right)}}{{P_g \lambda _m \left( {1 - e^{ - \lambda gt_0 } } \right)e^{ - \lambda gt_1 } \left( {1 - e^{ - \lambda gt_2 } } \right)}} \cdot \frac{{I_{\gamma m} }}{{I_{\gamma g} }} \cdot \frac{{\varepsilon _{\gamma m} }}{{\varepsilon _{\gamma g} }}, \label{f1}
\end{equation}
where $P_{m(g)}=\Phi \sigma_{m(g)} N$ is the number of radioactive nuclei produced by a certain reaction per unit of the exposition time, $\Phi$ is the beam flow, N is the number of nuclei in a  thin sample, $t_0$ -- exposure time, $t_1$ -- source holding time, $t_2$ is the $\gamma$-ray spectrum accumulation time, $\varepsilon _{\gamma}$ is the relative photoefficiency of the $\gamma$-line.

In the discussed scheme the isomeric ratio $(\sigma_m/\sigma_g)$ coincides with the relative yield of isomers ${P_m }/{P_g }$ so from formula (\ref{f1}), one can deduce the relationship

\begin{equation}
\frac{{\sigma _m }}{{\sigma _g }} = \frac{{S_m I_{\gamma g} \varepsilon _{\gamma g} \lambda _m \left( {1 - e^{ - \lambda gt_0 } } \right)e^{ - \lambda gt_1 } \left( {1 - e^{ - \lambda gt_2 } } \right)}}{{S_g I_{\gamma m} \varepsilon _{\gamma m} \lambda _g \left( {1 - e^{ - \lambda mt_0 } } \right)e^{ - \lambda mt_1 } \left( {1 - e^{ - \lambda mt_2 } } \right)}} \label{f2}
\end{equation}

If there is a $\gamma$-transition between isomers, then the contribution of this process to the lower state (it may be high-spin or low-spin one)  yield must be taken into account. As an example, in the case that the low-spin state is lower in energy, the calculation of the yield of the lower isomeric state $Y_{g}$ is performed by the use of formula presented in Refs. \cite{lu,kao,dick},

\begin{widetext}
\begin{equation}
\begin{array}{l}
 Y_g  = QP_m \left\{ {\frac{{\lambda _g }}{{\lambda _g  - \lambda _m }}\,\left( {1 - e^{ - \lambda mt_0 } } \right)\,e^{ - \lambda mt_1 } \,\frac{{\left( {1 - e^{ - \lambda mt_2 } } \right)}}{{\lambda _m }} -  \frac{{\lambda _m }}{{\lambda _g  - \lambda _m }}\,\left( {1 - e^{ - \lambda gt_0 } } \right)\,e^{ - \lambda gt_1 } \,\frac{{\left( {1 - e^{ - \lambda gt_2 } } \right)}}{{\lambda _g }}} \right\} +  \\ 
  + \frac{{P_g }}{{\lambda _g }}\left( {1 - e^{ - \lambda gt_1 } } \right)\,e^{ - \lambda gt_1 } \,\left( {1 - e^{ - \lambda gt_2 } } \right), \label{f3}\\ 
\end{array}
\end{equation}
\end{widetext}
where Q is the relative probability of the isomeric transition to the lower state of the pair due to the emission both the $\gamma$-quantum and the electron of internal conversion. The expression of the isomeric ratio in this case takes the following form \cite{va},
\begin{widetext}
\begin{equation}
\frac{{\sigma _m }}{{\sigma _g }} = \left[ {\frac{{\lambda _g \left( {1 - e^{ - \lambda mt_0 } } \right)\,e^{ - \lambda mt_1 } \,\left( {1 - e^{ - \lambda mt_2 } } \right)}}{{\lambda _m \left( {1 - e^{ - \lambda gt_0 } } \right)\,e^{ - \lambda gt_1 } \,\left( {1 - e^{ - \lambda gt_2 } } \right)}}\,\left( {\frac{{S_g I_{\gamma m} \varepsilon _{\gamma m} }}{{S_m I_{\gamma g} \varepsilon _{\gamma g} }} - Q\frac{{\lambda _g }}{{\lambda _g  - \lambda _m }}} \right) + Q\frac{{\lambda _g }}{{\lambda _g  - \lambda _m }}} \right]^{ - 1}.\label{f4} 
\end{equation}
\end{widetext}
Calculations  were carried out using formulas (\ref{f2}) and (\ref {f4}). The times of exposure, source holding, and collection of the $\gamma$-ray spectrum were strictly fixed. 

The targets were metal foils or powders enriched with the corresponding isotopes. The thickness of the metal targets is determined by weighing and does not exceed 3 $\div$ 5 mg/cm$^2$. If the target was prepared from a powdery substance, then the powder was applied in a uniform layer on a thin 10-$\mu$m lavsan substrate and fixed with an alcoholic shellac solution. The thickness of the target did not exceed 5 mg/cm$^2$. That excluded the need to introduce corrections for self-absorption of $\gamma$-quanta in the target itself.

The energy of the incident alpha particles was varied using degraders (aluminum foils). The degraders were located close to the target. The foils thickness was varied from 7 to 200 $\mu$m. This made it possible to reduce the energy of alpha particles from the maximum possible value obtained at the cyclotron -- 25 or 31 MeV -- to 5 MeV. The spread of the energy distribution of $\alpha$-particle beam in the energy range of 5 $\div$ 15 MeV and 20  $\div$ 31 MeV was respectively 200 and 500 keV.  The thickness of the foils was determined by weighing with an accuracy of 0.001 mg. The energy of alpha particles passing through a degrader of a given thickness was determined from the tabular data \cite{ander}. The correctness of finding the energy of alpha particles was confirmed by the coincidence of the experimental values of the energies of the reaction thresholds ($\alpha$,n); ($\alpha$,2n); ($\alpha$,3n) and ($\alpha$,p) with those obtained by calculation.

The value of the error of isomeric ratio $(\sigma_m/\sigma_g)$ was determined as the mean square of the systematic and statistical errors as it is recommended in Ref. \cite{jel}. Systematic errors consist of:

1. Errors in determining the energy of the bombarding particles accelerated by the cyclotron, which did not exceed 2\%;

2. Errors of no more than 0.5\% in determining the energy of particles that have passed through a degrader of a certain thickness (tabular data);

3. Errors in determining $\lambda_m$, $\lambda_g$, and $I_\gamma$ -- the yield of $\gamma$-quanta per 100 decays (tabular data). For particular isotopes, errors in determining these values range from 3 to 5\%.

4. Errors in determining the efficiency of the $\gamma$-spectrometer which are $\sim$ 5\%.

Statistical errors are associated with the accuracy of determining the area under the $\gamma$-line full-energy peak, which is determined by the total number of recorded pulses by the detector. We selected mainly single intense $\gamma$-lines in the spectrum, for which there was no overlap with neighboring ones. They are better separated from Compton distributions and it is easier to interpolate the background under the  full-energy peaks. Special measurements of the background were carried out and the shape of a single $^{137}$Cs line was analyzed. For each $\gamma$-line, the contribution of the Compton scattering of $\gamma$-quanta produced by high energy $\gamma$-transitions was estimated. For intense lines significantly exceeding the background, it is approximately constant along the entire width of the $\gamma$-line. Usually, it was found by linear interpolation between points lying outside $E_\gamma \pm 3\Delta E$, where $\Delta E$ is the line width at half maximum as it is recommended in Ref. \cite{jel2}. In the case of $\gamma$-lines in the energy range up to 500 keV, the separation of the background is hampered by a steeper fall in the Compton distribution.

When processing $\gamma$-spectra, the areas under the full-energy peaks, after subtracting the background, were determined by summing the pulses. The statistical error in the area of full-energy peaks for particular nuclei did not exceed 2 $\div$ 5\% at the maximum energy of the bombarding particles. It has not a pronounced tendency to increase with decreasing energy. 

It is important to note that in the presence of a transition between the states of an isomeric pair, even relatively small total, including both the systematic and statistical components, errors in measuring the areas ${S_m }$ and ${S_g }$ can lead to a sharp decrease in the accuracy of evaluating the isomeric ratios at some values of the parameters included in formula (\ref{f4}). Therefore, the optimization of the time parameters of each such measurement is a very important element of this procedure. In some cases, this circumstance narrows the energy range in which reliable results can be obtained. There are even examples of isomeric pairs that are practically inaccessible for the studies by the described method. The problems of optimizing the time parameters of such measurements are discussed in Ref. \cite{chsh}. 

In the case of reactions with $\alpha$-particles, isomers studied in the reaction ($\alpha$,n) are also produced by the reactions ($\alpha$,2n) and ($\alpha$,3n)  on impurity isotopes differing in mass more by 1 and 2 mass units from the main isotope, respectively. Sometimes this contribution is noticeable even at small ($\sim$ 1 $\div$ 3\%) amounts of impurity isotopes and must be taken into account. Therefore it is necessary to analyze the influence of the contributions from ($\alpha$,2n) and ($\alpha$,3n) reactions on the final result of the experimental ratio $(\sigma_m/\sigma_g)$ in the ($\alpha$,n) reaction at energies above the thresholds of reactions ($\alpha$,2n) and ($\alpha$,3n). The results of such analysis are discussed below in cases where it is required.

A series of measurements were carried out to obtain the value of each particular isomeric ratio. The number of irradiated targets could reach few tens. The values of the isomeric ratios presented in this work are the result of several such series of measurements carried out in different years, i. e. they were refined several times in connection with the appearance of other works concerning these isomers and their population or the refinement of spectroscopic data in the tables of isotopes, or other motivations. In repeated measurements of the areas under the full-energy peaks of the $\gamma$-lines at a certain energy of the bombarding particles, their values were well reproduced. The  root mean square deviation $\Delta  = \sqrt {\Sigma \left( {\sigma _m /\sigma _g  - \overline{\sigma _m /\sigma _g }} \right)^2 /n\left( {n - 1} \right)} ;\quad n > 1$ was analyzed within the framework of the method presented in Ref. \cite{jel}. The total result was considered reliable if the root mean square deviation thus obtained did not go beyond the errors estimated in the framework of the procedures described above.

Thus the presented technique made it possible to successfully measure of the ICSR for isomers with a lifetime of few minutes, including when the higher lying state is characterized by such a lifetime and there is the isomeric transition.

\section{The experimental data and the analysis}

\subsection{
The yield of isomers ($J_m=6^+, J_g=2^+$) in $^{41}$K($\alpha$,n)$^{44}$Sc reaction}

The yield of isomers in the reaction $^{41}$K($\alpha$,n)$^{44}$Sc have been measured in the $\alpha$-particle energy range from 15.0 to 31.0 MeV. The targets were irradiated by $\alpha$-particles with the current of about 2 $\mu$A. The beam particles were slowed down using aluminum foils. We used a $^{41}$KCl target with the thickness of 5 mg$\cdot$ cm$^{-2}$ and enrichment 95.8\%. The ICSR values were determined from the intensities of lines 271 keV (isomeric transition, $I^{rel}_\gamma$ = 86.6 \%, taking into account the internal conversion rate) and line from the spectrum of the daughter nucleus $^{44}$Ca 1157 keV ($I^{rel}_\gamma$ = 99.89 \%) taking into account the change in intensity during the decay process of each of the two states of the isomeric pair. 

The presence of the intense spectral lines, the relatively long lifetimes of the ground and isomeric states make the isomeric pair of the $^{44}$Sc nucleus a popular object of investigations. The ICSR values for it have been studied by different experimental groups in various reactions. In particular, the reactions with gamma-quanta -- $^{45}$Sc($\gamma$,n)$^{44}$Sc \cite{zhel},  protons -- $^{44}$Ca(p,n)$^{44}$Sc \cite{sach} and heavy ions -- $^{29}$Si($^{18}$O,p2n)$^{44}$Sc \cite{seab} were put to use.

Studies of the discussed reaction were also actively carried out \cite{mats,ril,mats2,keed,lev,izv99}. The relatively easy enrichment of the sample with isotope $^{41}$K provides additional convenience for studying it. In some of these works -- Refs.  \cite{mats,mats2,izv99} -- the results of measuring isomeric ratios are presented.

\begin{figure}
\includegraphics[width=1.15\linewidth]{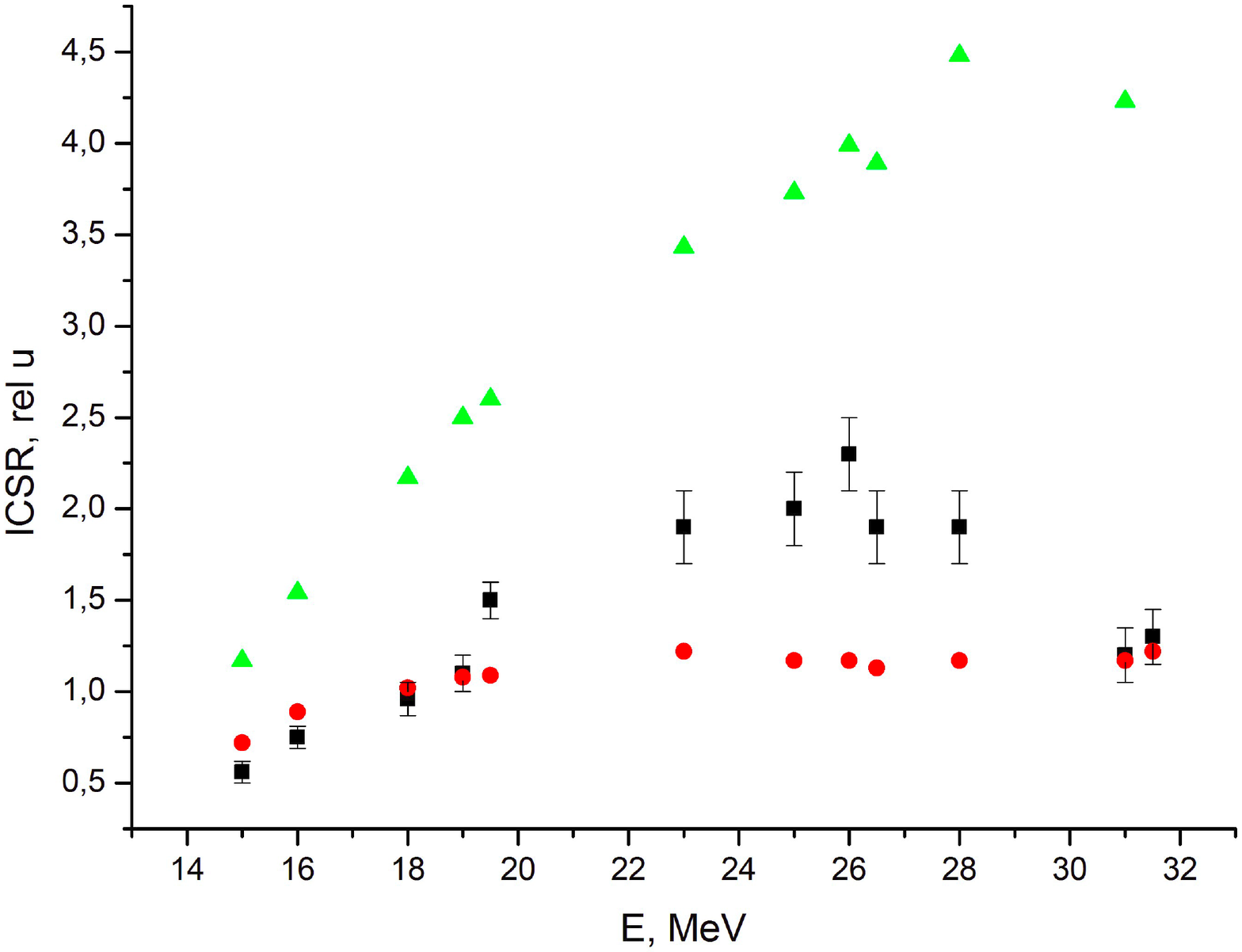} 
\caption{The ICSR of $^{44}$Sc obtained from the study of $^{41}$K($\alpha$,n) reaction (see comments in the text). \label{fig_41K}}
\end{figure}

The experimental results obtained by us, expanding and refining the data of work \cite{izv99}, are plotted in Tab. I together with the results of the earlier measurements presented in Ref. \cite{mats2}.  These data are in rather good agreement. Small differences take place at energies of 19.5, 23, and 28 MeV. The values obtained in our measurements seem to be more reliable, since it was obtained using a more sensitive technique, namely, the mentioned above Ge(Li) detector was used.  Data from both experiments are collected in Fig. \ref{fig_41K}.  For the cases when the energies coincide, the figure shows our data. The absolute values of the cross sections for the formation of the isomeric  $\sigma _m $ and ground  $\sigma_g$ states of the $^{44}$Sc nucleus are presented in Ref. \cite{mats2} and Ref. \cite{lev}. The values of both $\sigma _m $ and  $\sigma_g$ in Ref. \cite{lev} are approximately two times less than the ones presented in Ref. \cite{mats2}. Therefore, it is difficult to conclude about the magnitude of these cross sections. At the same time the values of isomeric ratios that can be deduced from the data of Ref. \cite{lev} are in good agreement with the data given in Tab. I.
 
\begin{center}
\begin{table}
\caption {Values of ICSR  $(\sigma _m /\sigma _g)$ and its aggregate errors $\Delta$ obtained from the study of the
reaction $^{41}$K($\alpha$,n)$^{44}$Sc.}
 \label{tab1}
\begin{tabular*}{0.42\textwidth}{ccccc}
\hline E (MeV) &  $(\sigma _m /\sigma _g)$ &$\Delta$ & $(\sigma _m /\sigma _g)$ \cite{mats2} & $\Delta$  \cite{mats2} \\
\hline
15.0& 0.56&0.06 & &\\
16.0& & & 0.75& 0.08\\
18.0&0.96 &0.09 & &\\
19.0& &&1.1 &0.1\\
19.5&1.5 &0.1 &1.2 &0.1\\
23.0&1.9 &0.2 &1.6 &0.1\\
25.0&& &2.0 &0.1\\
26.0&2.3 &0.2 & &\\
26.5& & &1.9 &0.2\\
28.0&1.9&0.2 &1.3 &0.1\\
31.0&1.2 &0.15 & &\\
31.5&& &1.3 &0.1\\
\hline
\end{tabular*}
\end{table}
\end{center}

It makes sense to pay attention to the values of ICSR obtained in the discussed experiments. They are many times superior to the ICSR observed in experiments with $\gamma$-quanta  \cite{zhel} and protons \cite{sach}. Moreover, even the ICSR values obtained from the study of the reaction induced by heavy ions, namely $^{29}$Si($^{18}$O,p2n)$^{44}$Sc \cite{seab} at higher energies of 30 $\div$ 99 MeV, are in the range of 1 $\div$ 5, that is, they are only slightly more than just presented despite the very large average moment brought by the high-energy $^{18}$O projectile.

Even more interesting is the fact that the curve of the dependence of A on the energy of $\alpha$-particles demonstrates a well-expressed maximum. In general, such a behavior of the measured quantities $(\sigma _m /\sigma _g)$ looks unexpected, because with an increase in the energy of $\alpha$-particles, the angular momentum brought into the compound nucleus also increases. Therefore, such a behavior is usually called anomalous. Pronounced peaks in the energy dependence of ICSR values similar to the one discussed here are rare, nevertheless some examples of this kind were observed by different groups. In spite of many calculations an interpretation of the maximum is an open question up to now.

Figure \ref{fig_41K} shows the results of calculations performed by use of the EMPIRE 3.1 \cite{he2} (triangles) and TALYS \cite{ko} (circles) codes too.  The former code  demonstrates a monotonic increase of ICSR and the calculated
values turn out to be much greater than the experimental ones.  At low (15 $ \div$ 19 MeV) and high ($\sim$ 30 MeV) energies, the code TALYS  gives a perfect description of the ICSR. However, at intermediate energies, calculations using the TALYS code do not reproduce a well-expressed experimental maximum.

\subsection{The yield of isomers ($J_m=9/2^+, J_g=1/2^-$) in $^{86}$Sr($\alpha$,n)$^{89}$Zr reaction}

Let us emphasize, once more, that in this and the next example we traditionally use the term "isomeric" ($m$) for states with high spin, although in the cases under discussion they are located below the low-spin states ($g$) of isomeric pairs. In addition, in this example, the activity of the nuclides that make up the isomeric pair is not independent, since they are linked by the isomeric transition itself. Because of this, the ICSR calculation is carried out using expression (\ref{f4}) with the transposition of the indices $g \leftrightarrow m$.

\begin{figure}
\includegraphics[width=1.15\linewidth]{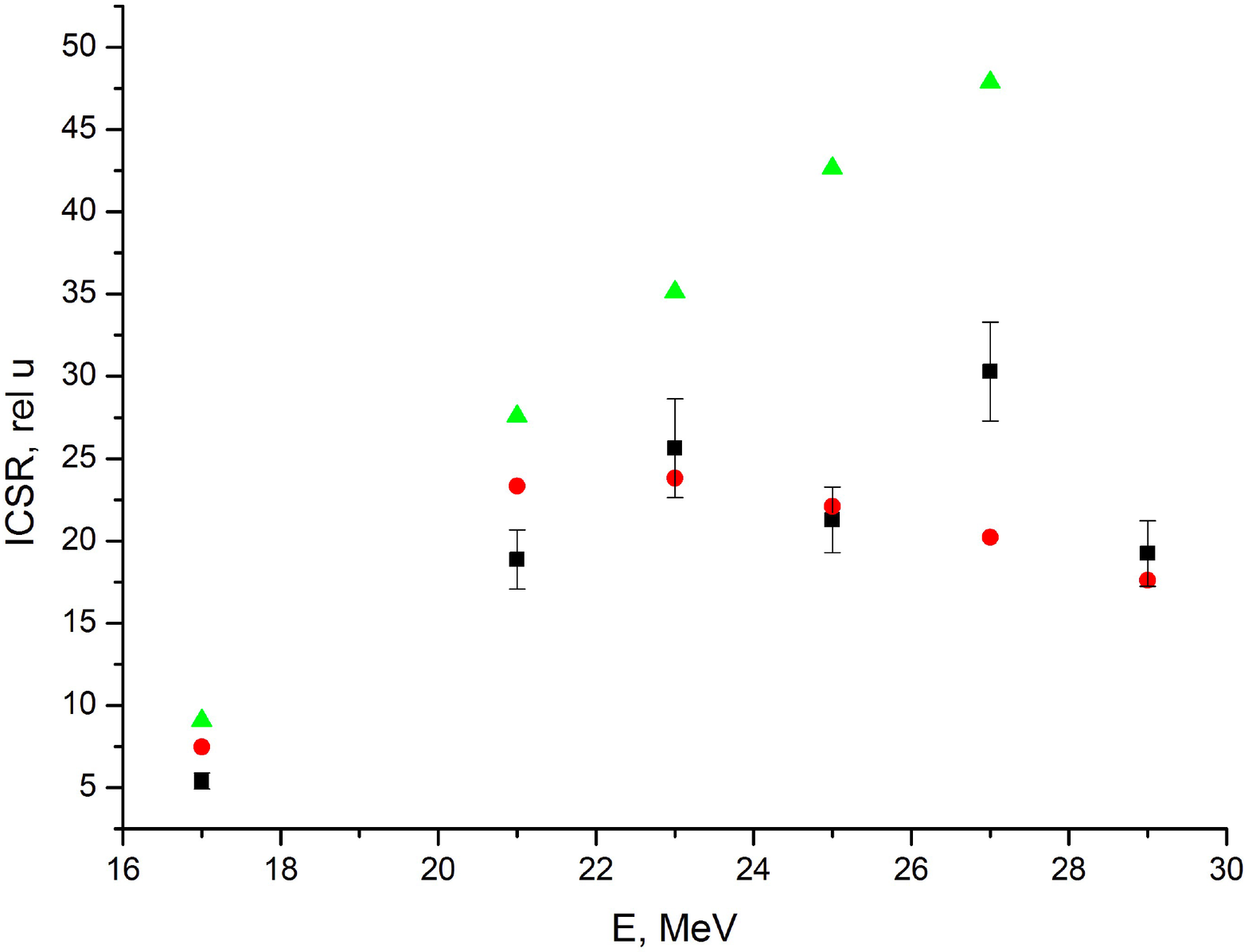} 
\caption{The ICSR of $^{89}$Zr obtained from the study of $^{86}$Sr($\alpha$,n) reaction (designations are the same as in Fig. \ref{fig_41K}). \label{fig_86Sr}}
\end{figure}

The  $^{89mg}$Zr isomers relative yield  data for the ($\alpha$,n) reaction were obtained at  $\alpha$-particle energies from 17 to 29 MeV. The ICSR values were determined from the measured intensities of the $\gamma$-line 588 keV (isomeric transition, $I^{rel}_\gamma$ = 93.8 \%) and the line from the spectrum of the daughter nucleus $^{89}$Y 909 keV ($I^{rel}_\gamma$ = 99.14 \%). The $^{86}$Sr target had impurities of $\sim$ 1.8\% $^{87}$Sr. Therefore, when studying the ICSR for the $^{86}$Sr($\alpha$,n) reaction, a contribution from the $^{87}$Sr($\alpha$,2n) reaction on impurities to the target is observed. The excitation function of the reaction ($\alpha$,2n) has a maximum value at an energy of 24 $\div$ 27 MeV, when the cross section ($\alpha$,n) of the reaction decreases significantly. In this case, when determining the isomeric ratio obtained from the study of $^{86}$Sr($\alpha$,n)$^{89}$Zr reaction, the contribution from the ($\alpha$,2n) reaction on $^{87}$Sr impurity was taken into account. The short lifetime of $1/2^-$ state (4.18 m.) 
creates a need for optimization of the times  $t_0$, $t_1$, and $t_2$. In the present paper we display the results, improved through the handling of the activation data with the use of the optimal extraction formula (\ref{f4}). As it is proved in \cite{chsh} both the use of this formula and the correct choice of timing parameters are very important, since ignoring this leads to extremely large errors in determining the ICSR. The experimental value of the ICSR produced by $^{86}$Sr($\alpha$,n)$^{89}$Zr reaction versus the $\alpha$-particle energy is plotted in Tab. II and shown in Fig.\ref{fig_86Sr}. 

\begin{center}
\begin{table}
\caption {Values of ICSR  $(\sigma _m /\sigma _g)$ and its aggregate errors $\Delta$ and $\delta$ obtained from the study of
reaction $^{86}$Sr($\alpha$,n)$^{89}$Zr.}
 \label{tab2}
\begin{tabular*}{0.47\textwidth}{ccccc}
\hline E (MeV) &  $(\sigma _m /\sigma _g)$ &$\Delta(\sigma _m /\sigma _g)$&$\sigma_{t}$, mb \cite {lev}&$\delta(\sigma_{t})$  \cite {lev} \\
\hline
17.0&5.4 & 0.5&670&10\%\\
21.0&18.9 &1.8 &270&10\%\\
23.0&25.6 & 3.0&160&10\%\\
25.0&21.3&2.0&97& 10\%\\
27.0&30.3 &3.0&63&10\% \\
29.0& 19.2&2.0&47&10\% \\
\hline
\end{tabular*}
\end{table}
\end{center}

The results of the experiment, in which the total cross section for the population of both isomeric states in the discussed reaction was measured \cite{lev}, are also accumulated in Tab. II. For these purposes the original excitation function of the reaction was interpolated 
in order to bring its values in line with the energies of the $\alpha$-particles used in our experiment. For all measured values in Tab. II and in Fig. \ref{fig_86Sr}, either absolute ($\Delta(\sigma _m /\sigma _g)$) or relative ($\delta(\sigma_{t})$) errors are given. Fig. \ref{fig_86Sr} shows in addition the results of calculations of the ICSR  in the framework of the EMPIRE 3.1 \cite{he2} and TALYS \cite{ko} codes.

Our experiments and those presented in \cite{lev}, demonstrate that both the relative yield of the high-spin isomer in the discussed reaction at  $\alpha$-particle energies 21 $\div$ 29 MeV and the total cross section for its population at  $\alpha$-particle energies 17 $\div$ 23 MeV are extremely large. Fig. \ref{fig_86Sr} shows that the results of the ICSR theoretical computations using the TALYS code reproduce well these results in the entire measurement range, and thus confirm this statement. 

\subsection{The yield of isomers ($J_m=7/2^+, J_g=1/2^+$) in $^{112}$Sn($\alpha$,n)$^{115}$Te reaction}

The yield of isomers in reaction $^{112}$Sn($\alpha$,n)$^{115}$Te have been measured in the $\alpha$-particle energy range from 20 to 30.4 MeV. The ICSR values were determined from the intensities of lines observed in the spectrum of daughter nucleus $^{115}$Sb, 770 keV (populated by the decay of low-spin isomeric state, $I^{rel}_\gamma$ $\sim$ 34.2 \%) and the group of lines 724 keV, 1327 keV, 1381 keV (populated by the decay of high-spin ground state with $I^{rel}_\gamma$ $\sim$ 30\%, $I^{rel}_\gamma$ $\sim$ 22.7\%, $I^{rel}_\gamma$ $\sim$ 23.0\% respectively). It was taken into account that $\gamma$-line 724 keV is populated in the decay of both discussed states. With an increase in the $\alpha$-particle energy, the reaction channel ($\alpha$,3n) opens and the formation of $^{115}$Te becomes possible on the $^{114}$Sn impurity to the target. In this case, the reaction $^{114}$Sn($\alpha$,3n)$^{115}$Te ($Q_r$ = --29.9 MeV) occurs, but since the target enriched in $^{112}$Sn contains $^{114}$Sn in an amount of 0.45\% and the reaction threshold is high enough, the contribution of this reaction to the formation of $^{115}$Te isomers is negligible small (less than measurement error). Alpha particle irradiation of other stable Sn isotopes with A $>$ 114  (impurities contained in the target) does not lead to the formation of $^{115}$Te.

However, it is possible, since the target enrichment in $^{112}$Sn was only 83.7\%, that reactions ($\alpha$,n), ($\alpha$,2n), ($\alpha$,3n), ($\alpha$,p) and ($\alpha$,pn), proceeding on impurity isotopes of Sn, produce nuclides whose $\gamma$-radiation is close in energy to the energies of $\gamma$-lines, from which the isomeric ratio is determined. The $\gamma$-background of these reaction products was analyzed basing on the tabular data in detail to make sure that the selected $\gamma$-lines refer only to isomers. As in the previous example, short isomeric states lifetimes (5.8 m. and 6.7 m. respectively) requires  optimization of the times  $t_0$, $t_1$, and $t_2$. The situation is somewhat simplified by the fact that there is no isomeric transition between states $J_m=1/2^+$ and $J_g=7/2^+$.

\begin{figure}
\includegraphics[width=1.15\linewidth]{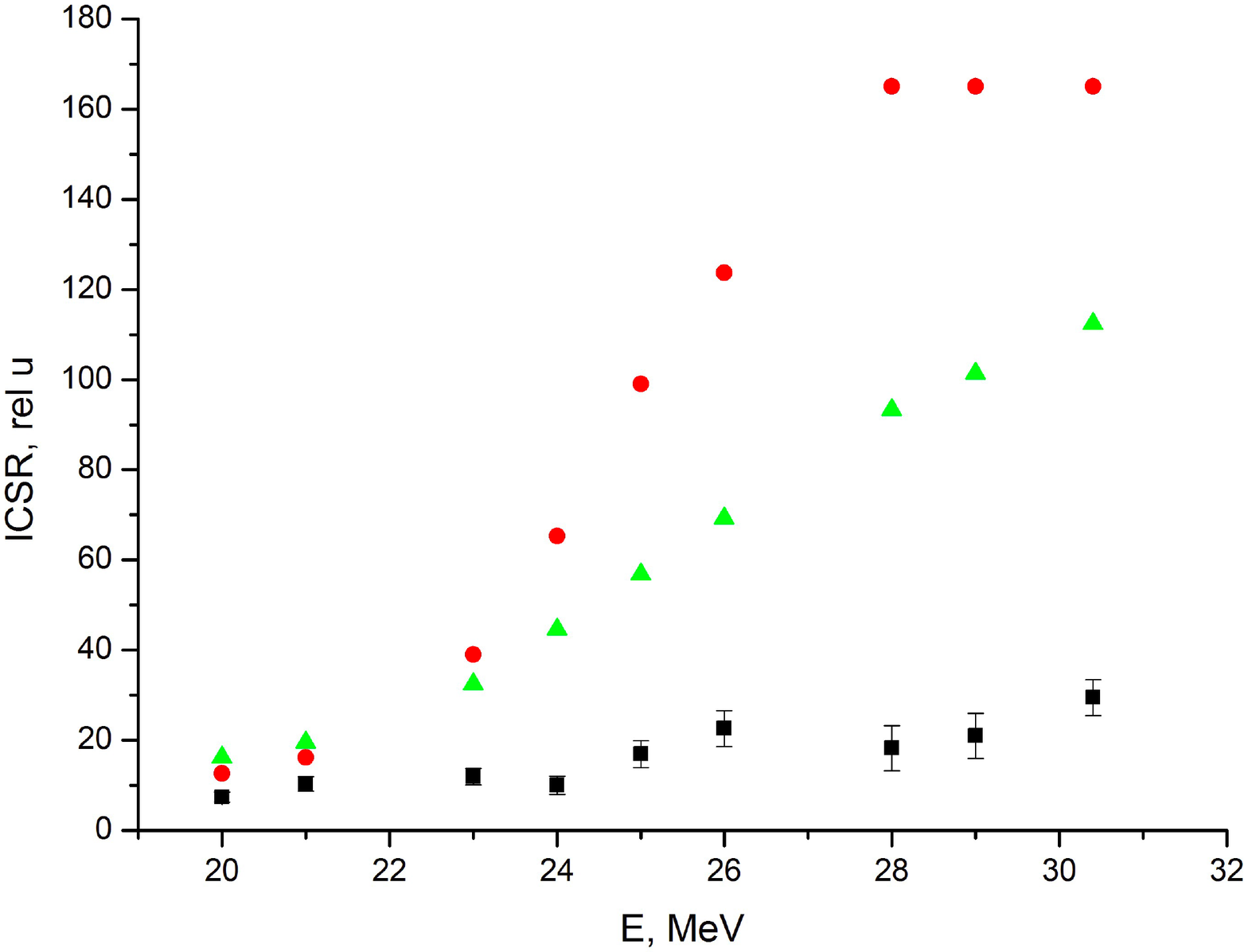} 
\caption{The ICSR of $^{115}$Te obtained from the study of $^{112}$Sn($\alpha$,n) reaction (designations are the same as in Fig. \ref{fig_41K}). \label{fig_112Sn}}
\end{figure}

The experimental results are plotted in Tab. III,  and are shown, together with the results of calculations performed by the use of the EMPIRE 3.1 and TALYS codes, in Fig. \ref{fig_112Sn}. The table  demonstrates that as is the case  with the ICSR of $^{86}$Sr($\alpha$,n)$^{89}$Zr reaction, great values of  $(\sigma _m /\sigma _g)$ are obtained. Nevertheless, both the EMPIRE 3.1 and the TALYS codes greatly overestimate even these large values, as indicated in Fig \ref{fig_112Sn}.  

What about the absolute values of the cross section for the formation of the $^{115}$Te isotope in this reaction, the energy range of the measured values ends at just 20 MeV \cite{khu}. The value of $\sigma_{t}$ is about 40 mb. in this point. 

\begin{center}
\begin{table}
\caption {Values of ICSR  $(\sigma _m /\sigma _g)$ and its aggregate errors $\Delta$ obtained from the study of reaction $^{112}$Sn($\alpha$,n)$^{115}$Te.}
 \label{tab1}
\begin{tabular*}{0.19\textwidth}{ccc}
\hline E (MeV) & $(\sigma _m /\sigma _g)$ &$\Delta$  \\
\hline
20.0&7.4 &1.1 \\
21.0&10.3 &1.6 \\
23.0&11.9 &1.8 \\
24.0&10.0 &2.0 \\
25.0&16.9 &3.0 \\
26.0&22.6 &4.0 \\
28.0&18.2 &5.0 \\
29.0&21.0 &5.0 \\
30.4&29.5 &4.0 \\
\hline
\end{tabular*}
\end{table}
\end{center}

\subsection{The yield of isomers ($J_m=11/2^-, J_g=3/2^+$) in $^{134}$Ba($\alpha$,n)$^{137}$Ce reaction}

The yield of isomers in reaction $^{134}$Ba($\alpha$,n)$^{137}$Ce  have been measured in the $\alpha$-particle energy range from 15.0 to 29.0 MeV. The targets were irradiated with $\alpha$-particles with the current of about 2 $\mu$A. The beam particles were slowed down using aluminum foils. We used a $^{134}$BaCO$_2$ target with the thickness of 5 mg$\cdot$ cm$^{-2}$ and enrichment 85.5\%. The ICSR values were determined from the intensity of line 254 keV (isomeric transition, $I^{rel}_\gamma$ = 11.3 \%, taking into account the internal conversion rate) and the aggregate intensity of two lines from the spectrum of the daughter nucleus $^{137}$La  447 keV  and 437 keV ($I^{rel}_\gamma$ = 2.1 \%) taking into account the change in intensity during the decay process of each of the two states of the isomeric pair. As in the previous examples, due to the limited target enrichment  the $\gamma$-background of the  products of ($\alpha$,2n) and ($\alpha$,3n) reactions occurring on impurity isotopes of Ba was analyzed in detail basing on the tabular data.  

The experimental results are plotted in Tab. IV together with the results of the measurements presented in Ref. \cite{sva}.  These data are in a good agreement. An exception is the point corresponding to an energy of 15 MeV. The value obtained by us seems to be more reliable, since it was obtained using a more sensitive technique, namely, the mentioned above Ge(Li) detector was used. In addition, in our measurements, we managed to find a point where the ICSR has a pronounced maximum. 

\begin{figure}
\includegraphics[width=1.15\linewidth]{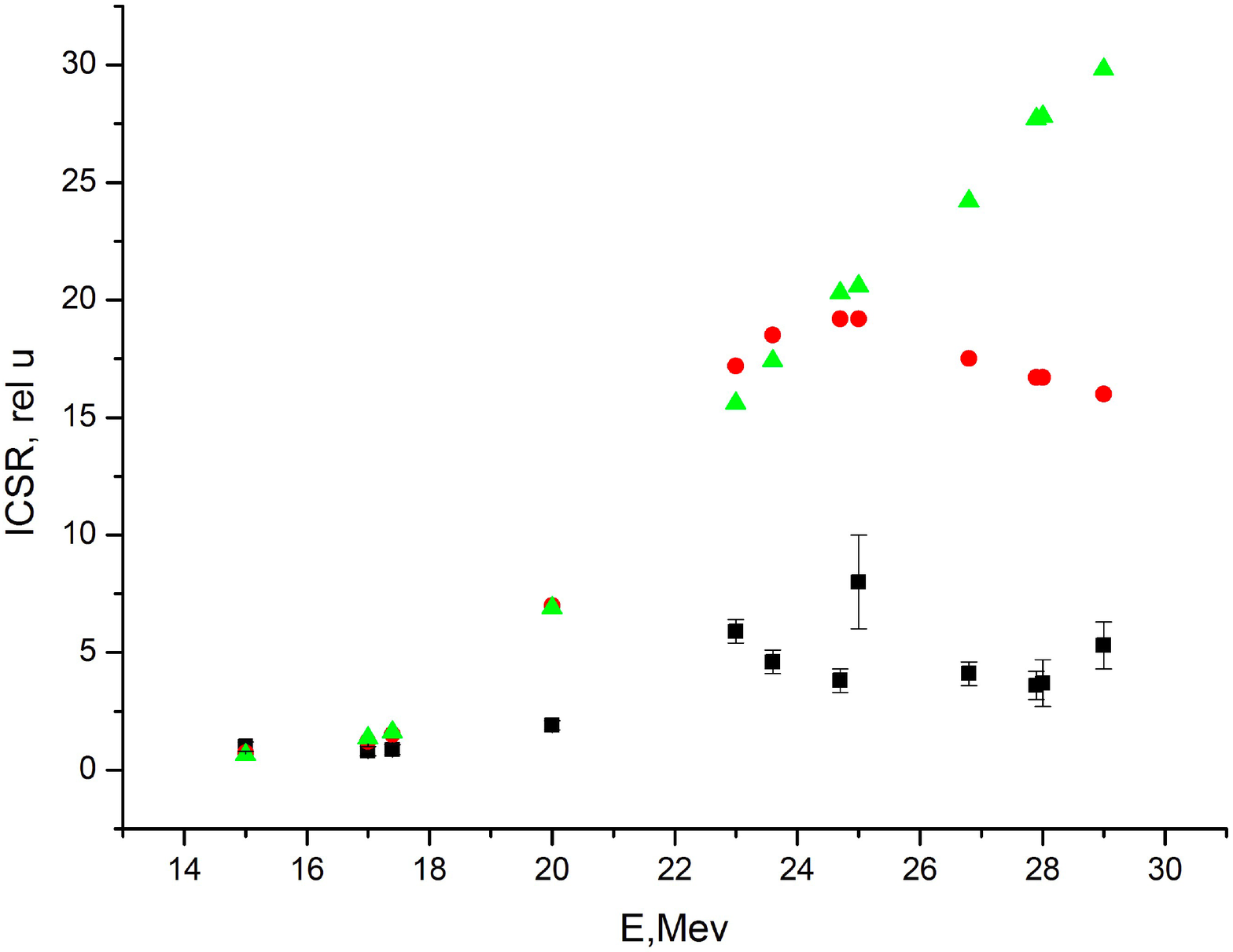} 
\caption{The ICSR of $^{137}$Ce obtained from the study of $^{134}$Ba($\alpha$,n) reaction  (designations are the same as in Fig. \ref{fig_41K}).\label{fig_134Ba}}
\end{figure}

Data from both experiments are collected in Fig. \ref{fig_134Ba}. The same figure shows the results of calculations performed by use of the EMPIRE 3.1 and TALYS codes. In the energy range 15 $\div$ 18 MeV, both the experimental results and the results of calculations obtained by the two methods actually coincide. At higher energies, both theoretical approaches give grately overestimated results.

\begin{center}
\begin{table}
\caption {Values of ICSR  $(\sigma _m /\sigma _g)$ and its aggregate errors $\Delta$ obtained from the study of
reaction $^{134}$Ba($\alpha$,n)$^{137}$Ce.}
 \label{tab4}
\begin{tabular*}{0.29\textwidth}{cccc}
\hline E (MeV) &  $(\sigma _m /\sigma _g)$ &$\Delta$  & $(\sigma _m /\sigma _g)$ \cite{sva}\\
\hline
15.0&1.0 & 0.2&0.43\\
17.0&0.80 &0.2 &\\
17.4&&&  0.87\\
20.0&1.9 &0.2 &1.9\\
23.0&5.9 & 0.5&\\
23.6&&&4.6\\
24.7&&&3.8 \\
25.0&8.0 &2.0 &\\
26.8&& &4.1 \\
27.9&& &3.6 \\
28.0&3.7 &1.0 &\\
29.0&5.3 & 1.0&\\
\hline
\end{tabular*}
\end{table}
\end{center}

\section{Discussions and conclusions}

In the present paper the results of the investigations of the yield of isomers in ($\alpha $,n) reactions namely  $^{41}$K($\alpha $,n)$^{44}$Sc, $^{86}$Sr($\alpha $,n)$^{89}$Zr, $^{112}$Sn($\alpha$,n)$^{115}$Te, and $^{134}$Ba($\alpha$,n)$^{137}$Ce in the energy range of the alpha particles E=15 $\div$ 31 MeV are presented. Much of these results were obtained in the framework of improved activation technique allows one to work with short-lived samples with coherent activity. The results substantially supplement the data on the yield of isomers in reactions with $\alpha$-particles. On the other hand, they add complementary touches to the variegated pattern of the effects observed in the study of processes of this type and corresponding calculated results.

In the framework  of the study of reaction   $^{41}$K($\alpha $,n)$^{44}$Sc the current work has reliably confirmed the well-pronounced maximum in the energy dependence of the ICSR, presented earlier in  Refs.  \cite{mats2,izv99}. The position of this maximum and the value of ICSR in it are adjusted. This behavior of the ICSR, namely its sharp decrease with increasing energy and, therefore, with increasing the angular momentum of the $\alpha$-particle at one and the same impact parameter beginning from the point of maximum looks anomalous. Pronounced anomalies of such a type are quite rare. In the discussed energy region they are observed in ($\alpha$,n)-reactions $^{113}$In($\alpha$,n)$^{116}$Sb \cite{ch2} and $^{115}$In($\alpha$,n)$^{118}$Sb \cite{ba2}. A sharp maximum was also found in reaction $^{109}$Ag($\alpha$,3n)$^{110}$In  \cite{vi0} at an energy of about 55 $\div$ 60 MeV, the value of ICSR in it reaches 10 and rapidly drops down by about 2 times with increasing energy. The data concerning the left slope of the maximum up to an energy of 54 MeV were confirmed by Ref. \cite{was0}. Peaks  in the energy dependence of ICSR are also observed by us in reactions $^{86}$Sr($\alpha $,n)$^{89}$Zr and $^{134}$Ba($\alpha$,n)$^{137}$Ce, but they are not so expressive.

It is not always possible to reproduce the anomalous behavior of isomeric ratios with increasing energy in calculations, even with the use of advanced codes. Indeed, the code EMPIRE 3.1 demonstrates, with rare exceptions, a monotonic growth of ICSR, so any maxima of the energy dependence and saturation effects are not described within its framework. The code TALYS in some cases reproduces qualitatively, and in special cases, as it is seen from Fig. \ref{fig_86Sr}, quantitatively saturation pattern and weak peaks of ICSR. Obtained and confirmed in a series of experiments carried out by different groups in the reaction  $^{41}$K($\alpha $,n)$^{44}$Sc pronounced maximum cannot be reproduced.

Reactions $^{86}$Sr($\alpha $,n)$^{89}$Zr and $^{112}$Sn($\alpha$,n)$^{115}$Te are characterized by very large, reaching 30, values of the ICSR. These are the largest known values of ICSR for ($\alpha$,n)-reactions, although slightly smaller values ($\sim$20) are presented in the literature. A good example can be found in Ref. \cite{bas0} (these results are presented in {\it janis34} database), in which reaction  $^{180}$W($\alpha$,n)$^{183}$Os gives the ICSR values up to 17. The sole example of ICSR values  in this energy range which are even higher (up to 37) was obtained in works \cite{ba4,ba5} for reaction $^{107}$Ag ($\alpha$,2n)$^{109}$In. 
The great values presented here are good confirmation of the qualitative considerations presented in section 2.

It is important to point out that all the examples discussed in this paragraph, characterized by the large ICSR values, correspond to the situation when the high-spin isomeric state is located lower than the low-spin one. In this situation, large ICSR values are qualitatively explained by the fact that a deep minimum of the yrast band, corresponding to the high-spin state, encounters on the path of $\gamma$-cascades going along or close to this band.   As a result, the branching ratio of the transition to the high-spin state and its bypass on the way to the low-spin state turn out to be greater compare to the opposite case, when the low-spin state is located below the high-spin state. The discussed minimum is called "yrast trap". It is clearly visible on the left-hand side of Fig. 1. However, it cannot be argued that the situation when the high-spin isomeric state is below the low-spin state necessarily leads to large values of ICSR. Indeed, the maximum value of ICSR for reaction $^{176}$Hf($\alpha$,n)$^{179}$W does not exceed 2.4 \cite{par} and there are enough examples of such a type.

Obviously, the most interesting for fundamental research are processes in which long-lived excited rather than ground states of such a type are populated with a high probability. It is these processes that make it possible to obtain high-purity beams of metastable isomers for studying, for example, super-elastic processes, etc. It is obvious that reactions with $\alpha$-particles can give only high yields of high-spin metastable nuclei. So of greatest interest are the cases when the high-spin isomer is located above the low-spin one, and  at the same time the isomeric ratio is large. From this point of view, reaction $^{134}$Ba($\alpha$,n)$^{137}$Ce studied in the present work, in which the value of ICSR at an energy of 23 $\div$ 25 MeV reaches 5 $\div$ 8, seems to be promising. Another promising process was found by our group \cite{gl1}. Reaction $^{136}$Ce($\alpha$,n)$^{139}$Nd was considered, the value of ICSR reaches 10. The formation of beams of high-spin $^{137}$Ce and $^{139}$Nd isomers obtained on the primary beam of $\alpha$-particles in the frame of the SPIRAL ISOLDE scheme, seems to be quite realistic because the lifetime of these nuclides is many hours.

Energy dependence of ICSR values is significantly more sensitive to the features of the processes occurring at the moment of collision, as well as after it in the compound and residual nucleus compared to the total excitation functions of nuclear reactions and angular distributions of primary reaction products. Indeed, usually the yield of the nuclear states with very different spins strongly depends on particle capture and emission strength functions, the densities of high- and low-spin states in the high-energy part of the spectrum of a residual nucleus, discrete spectrum of this nucleus, probabilities of $\gamma$-transitions between discrete levels, etc. The advantage of the activation method in comparison with the ones in which particles directly generated by the primary collision are registered is that the discussed method makes it possible to reliably identify the mass of the residual nucleus and thus the reaction itself, in which it is produced, at energies far from the regions where these reactions are dominant. That is why the ICSR measurements seem to be a good test of applicability of nuclear reaction computer codes in a wide range of energies. The experimental results presented in this work make it possible to compare the quality of their description by the modern computer programs, in particular the EMPIRE 3.1 and TALYS. The code TALYS gives a rather good description of the yield of isomers in ($\alpha$,n)-reactions on relatively light nuclei $^{41}$K and  $^{86}$Sr although it does not reproduce the pronounced maximum observed in the former reaction. For heavier target nuclei $^{112}$Sn and  $^{134}$Ba this code qualitatively reproduces the saturation of the ICSR values at the $\alpha$-particle energy several MeV higher than the potential barrier height, but greatly overestimates the actual ICSR values. As noted above, the code EMPIRE 3.1 demonstrates a monotonic growth of ICSR and so does not describe saturation effects. Like the TALYS code, it also grately overestimates the ICSR values in the energy range 20$\div$30 MeV but for reaction $^{112}$Sn($\alpha$,n)$^{115}$Te this excess is somewhat less. Both codes reproduce well the ICSR values obtained in reaction  $^{134}$Ba($\alpha$,n)$^{137}$Ce at low energies.

Analysis of a wider list of ($\alpha$,n) reactions leads to approximately the same conclusions. The TALYS code has some advantages over the EMPIRE 3.1 code as it reproduces a qualitative picture of the saturation of ICSR values. Both codes give quantitative values of ICSR that differ significantly from the experimental ones often overestimating, but sometimes underestimating actual values. This applies especially to the energy range E$\geq 20$ MeV. There are also examples of a successful quantitative description. So, the EMPIRE 3.1 code describes well the ICSR values obtained in reaction $^{136}$Ce($\alpha$,n)$^{139}$Nd \cite{gl1}, and the TALYS code -- the ones obtained in reaction $^{187}$Re($\alpha$,n)$^{190}$Ir  \cite{ch10}. 

The origins for the discrepancy between the mentioned ICSR experimental data and results of various calculations (sometimes dramatic, as in some of the examples presented above) as well as the origins of occurring, although not often, cases of a successful description of the measurement results  are not disclosed yet. No systematic dependence of ICSR values on the spins of the populated isomeric states, their energies, the proximity of magic numbers, shape and other features of the atomic nucleus, etc. is found. These circumstances make a particularly strong impression in view of the fact that the total cross sections and angular distributions of various nuclear reactions induced by nucleons and light nuclear particles in most cases are described by modern computer codes, such as EMPIRE 3.1 and TALYS, quite satisfactorily. Taking into account the fact just mentioned, the most likely source of the described difficulties is the lack of information about the $\gamma$-cascades occurring in the residual nucleus (insufficient information on spin dependence of the level density of the nucleus,  absence of data on the transitions between rotational bands, etc.), although there are probably other sources as well. This problem could possibly be resolved by direct measurements of the characteristics of gamma cascades ending in isomeric states using a gamma sphere.

The results of the performed measurements, as well as the results of the previous ones, seem to some extent a challenge for the theory of nuclear reactions at low energies. Elimination of existing discrepancies between the experiment and the results of theoretical descriptions can shed light on the features of the angular momentum dynamics characteristic for these reactions.

So, the presented data and the results of other works used in the discussion demonstrate the importance and broad prospects of experimental and theoretical studies of the isomeric yields.

\end{document}